\documentclass{aa}
\usepackage{psfig}

\def\bron{GS~1826-238}
\def\ecs{erg~cm$^{-2}$s$^{-1}$}
\def\lum{erg~s$^{-1}$}

\begin{document}
\thesaurus{05(08.09.2 GS~1826-238; 08.14.1; 13.25.1; 13.25.5)}

\title{Broad-band X-ray measurements of \bron}
\titlerunning{Broad-band X-ray measurements of \bron}
\author{J.J.M.~in~'t~Zand\inst{1} 
 \and J.~Heise\inst{1}
 \and E.~Kuulkers\inst{1,2}
 \and A.~Bazzano\inst{3}
 \and M.~Cocchi\inst{3}
 \and P.~Ubertini\inst{3}}
\offprints{J.J.M.~in~'t Zand (e-mail {\tt jeanz@sron.nl})}

\institute{     Space Research Organization Netherlands, Sorbonnelaan 2,
                NL - 3584 CA Utrecht, the Netherlands
         \and
                Astronomical Institute, Utrecht University,
                P.O.Box 80000, NL - 3508 TA Utrecht, the Netherlands
         \and
                Istituto di Astrofisica Spaziale (CNR), Area Ricerca Roma Tor
                Vergata, Via del Fosso del Cavaliere, I - 00133 Roma, Italy
                        }
\date{Received 12 February 1999, accepted 6 April 1999}

\maketitle

\begin{abstract}
The broad band X-ray spectrum of the low-mass X-ray binary (LMXB) \bron\ was 
measured with the narrow-field instruments on {\em BeppoSAX\/} on April 6 and 
7, 1997. The spectrum is consistent with the Comptonization of a 0.6~keV 
thermal spectrum by a hot cloud of temperature equivalent $kT=20$~keV. 
During the observation two type I X-ray bursts were detected. From the bursts 
an upper limit to the distance could be derived of 8~kpc. Combined with an
elsewhere determined lower limit of 4~kpc this implies a persistent X-ray 
luminosity between $3.5\times10^{36}$ and $1.4\times10^{37}$~\lum\ which is 
fairly typical for a LMXB X-ray burster.
The accurate determination of the energetics of the two bursts and the 
persistent emission confirm results with the Wide Field Cameras on 
{\em BeppoSAX\/} in a narrower bandpass (Ubertini et al. 1999).
Comparison with independent X-ray measurements taken at other times indicates
that \bron\ since its turn-on in 1988 is a rather stable accretor, which is in 
line with the strong regularity of type I X-ray bursts.
\end{abstract}

\section{Introduction}
\label{intro}

A decade after its discovery (Makino et al. 1989), the nature of the compact 
object in the X-ray binary \bron\ has finally been established. Monitoring 
observations with the Wide Field Cameras (WFCs) on {\em BeppoSAX\/} revealed the source 
to be a regular source of type I X-ray bursts which are explained as 
thermonuclear runaway processes on the hard surface of a neutron star (Ubertini 
et al. 1997, 1999). Previously, the nature was under debate because the X-ray
emission exhibited characteristics that were until recently suspected to be 
solely due to black hole candidates (Tanaka 1989).

An optical counterpart has been identified (Motch et al. 1994, Barret et al.
1995) which classifies the binary as a low-mass X-ray binary (LMXB). 
Later this counterpart was found to exhibit optical bursts and a modulation 
which is likely to have a periodicity of 2.1~h (Homer et al.
1998). If the latter is interpreted to be of orbital origin, it would imply a 
binary that is compact among LMXBs.

\bron\ appears unusual among LMXBs. 
First, X-ray flux measurements after its 1989 discovery are fairly
constant (In~'t~Zand 1992, Barret et al. 1995) at a level 
of approximately $6\times10^{-10}$~\ecs\ in 2 to 10 keV. 
Second, the WFC measurements reveal a strong regularity in the occurrence
of type I X-ray bursts for an unusually long time (Ubertini 
et al. 1999). These two facts are very probably related. The constant flux
is indicative of a stable accretion of matter on the neutron star which
fuels regularly ignited thermonuclear explosions that give rise to X-ray bursts.

\bron\ has a hard spectrum, the initial {\em Ginga\/} observations measured
a power law spectrum with a photon index of 1.8 (Tanaka 1989). This makes 
it particularly important to study the spectrum in a broad photon energy
range. Strickman et al. (1996) have 
attempted this by combining the early {\em Ginga} 1-40 keV data with 60-300 
keV OSSE data taken in 1994. 
Del Sordo et al. (1998) have performed a preliminary study of the 0.1-100
keV data taken with the narrow-field instruments (NFI) on board {\em BeppoSAX} in 
October 1997. In the present paper, we study data taken with the same 
instrumentation half a year before that. The primary purpose of this study is 
to accurately analyze the flux of the persistent emission as well as that of 
two X-ray bursts. Also, we study the variability of the 2 to 10 keV emission.

\begin{figure}[t]
\psfig{figure=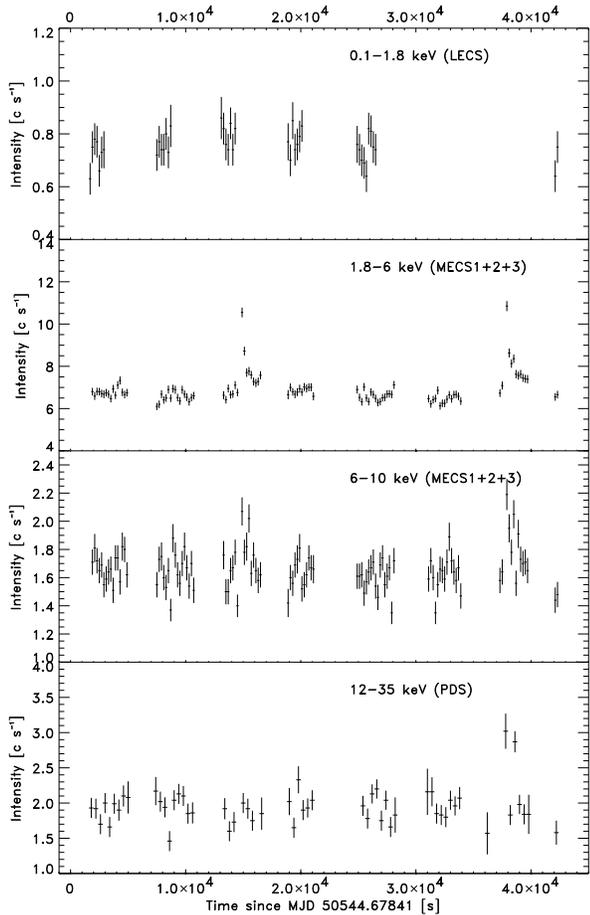,width=\columnwidth,clip=t}

\caption[]{Light curves of \bron\ persistent emission. Background contributions
have been subtracted. The data during the burst time intervals from -7 to 
+113~s with respect to the burst peak times have been excluded (see
Figs~\ref{figburstlc1} and \ref{figburstlc2}). The time 
resolution is 200~s for the LECS and MECS data and 400~s for the PDS data.
\label{figpersistentlc}}
\end{figure}

\section{Observations}
\label{secspec}

The NFI include the Low-Energy and the Medium-Energy Concentrator Spectrometer 
(LECS and MECS, see Parmar et al. 1997 and Boella et al. 1997 respectively) 
with effective bandpasses of 0.1-10 and 1.8-10~keV, respectively. Both are 
imaging instruments. The MECS was used in the complete configuration of three 
units (unit 1 failed one month after the present observation).
The other two NFI are the Phoswich Detector System (PDS;
active between $\sim12$ and 300 keV; Frontera et al. 1997) 
and the High-Pressure Gas Scintillation Proportional Counter (HP-GSPC;
active between 4 and 120 keV; Manzo et al. 1997). 

A target-of-opportunity observation (TOO) was performed with the NFI between
April 6.7 and 7.2, 1997 UT (i.e., 40.8~ks time span). The trigger for the TOO
was the first recognition that the source was bursting (Ubertini et al. 1997). 
The net exposure times are 8.2~ks
for LECS, 23.1~ks for MECS, 18.0~ks for HP-GSPC and 20.5~ks for PDS. 
\bron\ was strongly detected in all instruments and two $\sim$150~s 
long X-ray bursts were observed.

We applied extraction radii of 8\arcmin\ 
and 4\arcmin\ for photons from LECS and MECS images, encircling at least 
$\sim95$\% of the power of the instrumental point spread function, to obtain 
lightcurves and
spectra. Long archival exposures on empty sky fields were used to define the
background in the same extraction regions. These are standard data sets
made available especially for the purpose of background determination. 
All spectra are rebinned so as to
sample the spectral full-width at half-maximum resolution by three bins
and to accumulate at least 20 photons per bin. The latter will ensure the
applicability of $\chi^2$ fitting procedures. A systematic error of 1\% is
added to each channel of the rebinned LECS and MECS spectra, to account
for residual systematic uncertainties in the detector calibrations 
(e.g., Guainazzi et al. 1998). For spectral analyses, the bandpasses were 
limited to 0.1--4.0 keV
(LECS), 2.2--10.5~keV (MECS), 4.0--30.0 keV (HP-GSPC) and 15--200 keV (PDS)
to avoid photon energies where the spectral calibration of the instruments
is not yet complete. In spectral modeling, an allowance was made to
leave free the relative normalization of the spectra from LECS, PDS and HP-GSPC
to that of the MECS spectrum, to accommodate cross-calibration uncertainties
in this respect. Use was made of the publicly available response matrices
(version September 1997).

\section{The persistent emission}
\label{secsppe}

\begin{figure}[t]
\psfig{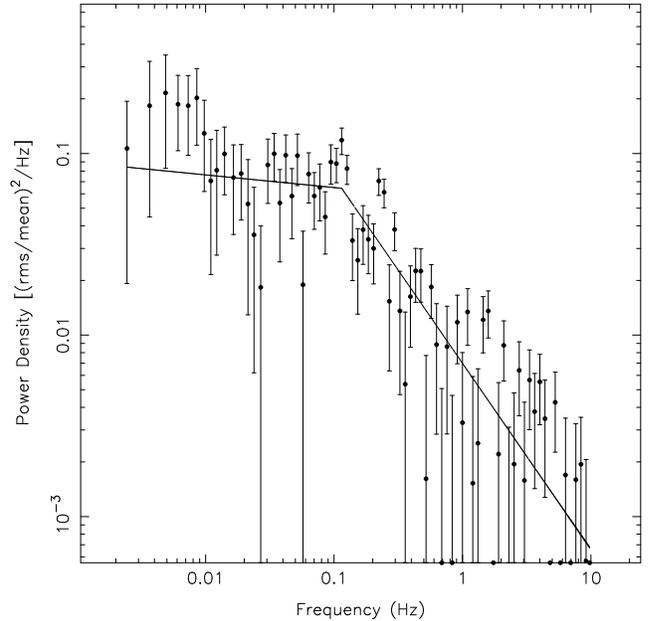}

\caption[]{Fourier power density spectrum of the intensity time series as 
measured with MECS between 2 and 10~keV. This is the average of the power
spectra of 31 time intervals of 819.2~s each with a time resolution of 0.05~s.
The Nyquist frequency is 10~Hz and the original frequency resolution 1/819.2~Hz.
The frequency domain was logarithmically rebinned. The drawn line shows a broken 
power law function fitted to the data.
\label{figpower}}
\end{figure}

Fig.~\ref{figpersistentlc} shows the lightcurve of the persistent
emission of \bron\ in various bandpasses. On time scales of a few hundred 
seconds, the flux appears constant except for immediately after the occurrence
of the two bursts. We searched for a modulation on time scales of about
2.1~h in the MECS data (which are the most sensitive) and find none. The
$3\sigma$ upper limit on the semi amplitude is 1.6\%. Thus, we cannot,
in X-rays, confirm the optical modulation which had a semi amplitude 
of 6\%. The power spectrum of the same data (excluding the burst intervals) is 
shown in Fig.~\ref{figpower}. A broken power law function was fitted to these 
data. The Poisson level, which is a free parameter, has been subtracted in 
Fig.~\ref{figpower}. Formally, the fit is unacceptable
($\chi^2=134$ for 72 dof). This may be due to narrow features at 0.2-0.3~Hz
and 1-2~Hz but the statistical quality of the data do not allow a detailed 
study of those. The break frequency of the broken power law is 
$0.115\pm0.011$~Hz, the power law index is $-0.07\pm0.10$ below and 
$-1.02\pm0.12$ above the break frequency. The high-frequency index is 
consistent with the index found from {\em Ginga\/} data taken in 1988 
between 0.1 and 500~Hz (Tanaka 1989). The integrated rms power of the noise 
between 0.002 and 10~Hz is $20\pm2$\%, that for the {\em Ginga\/} data 
between 0.02 and 500 Hz is $\sim30$\% (Barret et al. 1995). 
If one assumes the same break frequency for the {\em Ginga}
data, we expect an rms of 17\% for these between 0.002 and 10~Hz which
is very similar to the MECS result. The break 
frequency is comparable with values found for LMXB atoll sources and is one order of magnitude below values 
found for the bright LMXB Z sources (see Wijnands \& Van der Klis 1999). 
We are unable to assess the history or variability of the low-frequency 
index or break frequency.

\begin{figure}[t]
\psfig{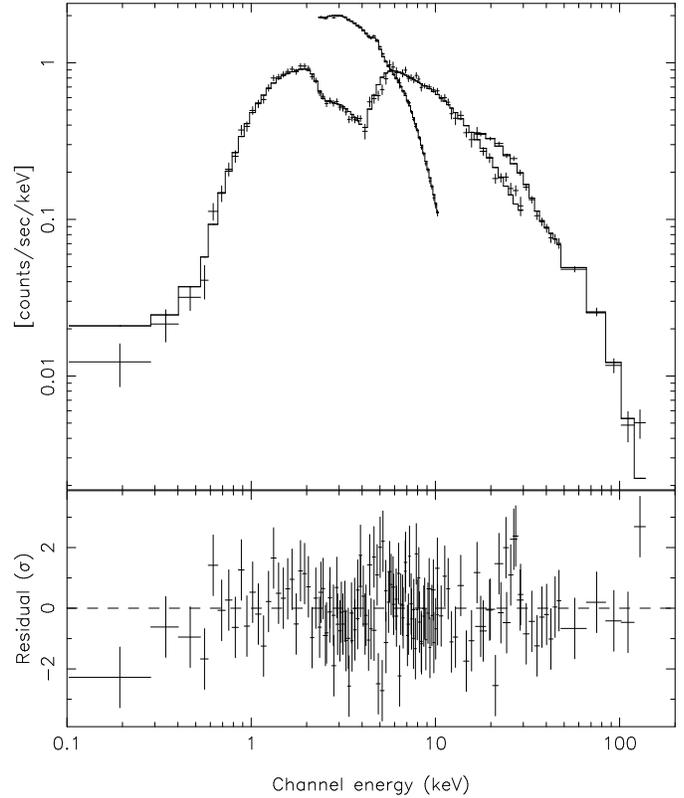}

\caption[]{Upper panel: count rate spectrum (crosses) and Comptonized spectrum
model (histogram) for average persistent emission (excluding -7/+1113~s time
intervals around the burst peaks). Lower panel: residual in units
of $\sigma$ per channel.
\label{figpersistentspectrum}}
\end{figure}

\begin{table}[t]
\caption[]{Parameter values of two spectral models fitted to the persistent 
emission. The errors are single parameter $1\sigma$ errors.}
\begin{tabular}{ll}
\hline
Model              & Comptonized spectrum plus black body\\
                   & ({\tt XSPEC: wa constant bb comptt})\\
$N_{\rm H}$        & $(1.1\pm0.2)\times10^{21}$ cm$^{-2}$ \\
bb $kT$            & 3.78$\pm0.32$ keV\\
bb $R$             & $0.21 \pm 0.03~d_{\rm 10~kpc}$ km \\
Wien $kT_{\rm W}$  & $0.581\pm0.010$ keV \\
Plasma $kT_{\rm e}$& $20.4\pm1.1$ keV\\
Plasma optical           & $2.13\pm0.10$ for disk geometry \\
\hspace{2mm}depth $\tau$ & $4.95\pm0.21$ for spherical geometry \\
Comptonization           & $0.72\pm0.07$ for disk geometry \\
\hspace{2mm}parameter $y$& $3.91\pm0.33$ for spherical geometry\\
$\chi^2_{\rm r}$   & 1.225 (132 dof) \\
Flux 2-10 keV      & $5.42\times10^{-10}$~\ecs \\
Flux 0.1-200 keV   & $1.93\times10^{-9}$~\ecs \\
\hline
Model              & cut off power law plus black body\\
                   & ({\tt XSPEC: wa constant bb cutoffpl})\\
$N_{\rm H}$        & $(5.4\pm0.2)\times10^{21}$ cm$^{-2}$ \\
bb $kT$            & 0.91$\pm0.03$ keV\\
bb $R$             & $3.1 \pm 0.3~d_{\rm 10~kpc}$ km \\
Photon index       & $1.38\pm0.03$ \\
Cut off            & $51.69\pm0.03$ keV \\
$\chi^2_{\rm r}$   & 1.534 (137 dof) \\
Flux 2-10 keV      & $5.41\times10^{-10}$~\ecs \\
Flux 0.1-200 keV   & $2.00\times10^{-9}$~\ecs \\
\hline
\end{tabular}
\label{tabpersistentspectrum}
\end{table}

A broad-band spectrum was accumulated, averaged over the complete observation
except the burst intervals, making use of the LECS, MECS, HP-GSPC, and PDS 
data. The spectrum was fitted with two models: black body radiation with 
unsaturated Comptonization (Titarchuk 1994), this is a model which is rather 
successful in describing 
other low-luminosity LMXBs as well (e.g., Guainazzi et al. 1998, In 't Zand et
al. 1999); and black body radiation plus a power law component with an
exponential cut off, this model was used by Del Sordo et al.
(1998) for NFI data below 100~keV on \bron. The results are given in 
Table~\ref{tabpersistentspectrum}, a graph is shown in 
Fig.~\ref{figpersistentspectrum} for the Comptonized model.

\begin{figure}[t]
\psfig{figure=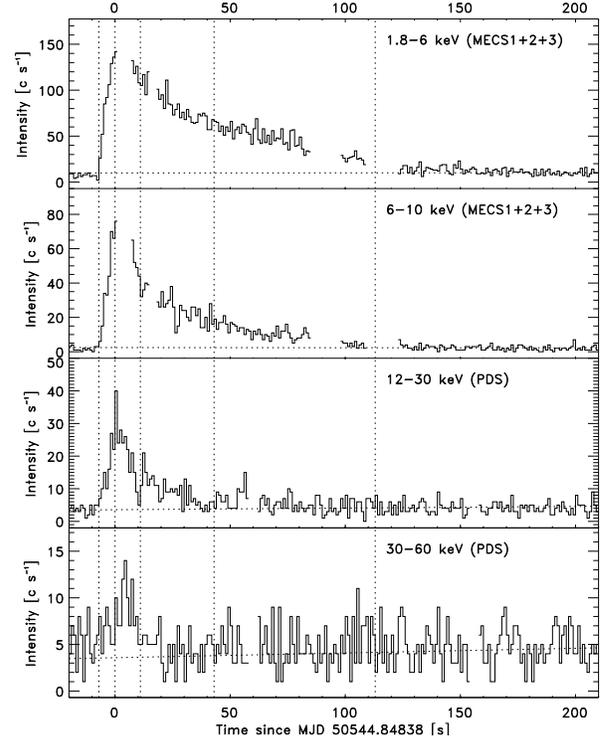,width=\columnwidth,clip=t}

\caption[]{Time profile of the first burst detected from \bron\ with MECS and 
PDS. The time resolution is 1~s. The curves have not been corrected for
contributions from the background and the persistent emission. The gaps
in the MECS data correspond to times of bad data.
The PDS curves were generated from collimator A and collimator B data, the
gaps correspond to collimator slew times between off and on source position
(these last 3~s and occur every 90~s). The linear trends as determined from 
the initial and final 10~s are drawn as near-horizontal dashed lines. The 
vertical lines indicate borders of time intervals for which spectra were 
modeled. 
\label{figburstlc1}}
\end{figure}

\begin{figure}[t]
\psfig{figure=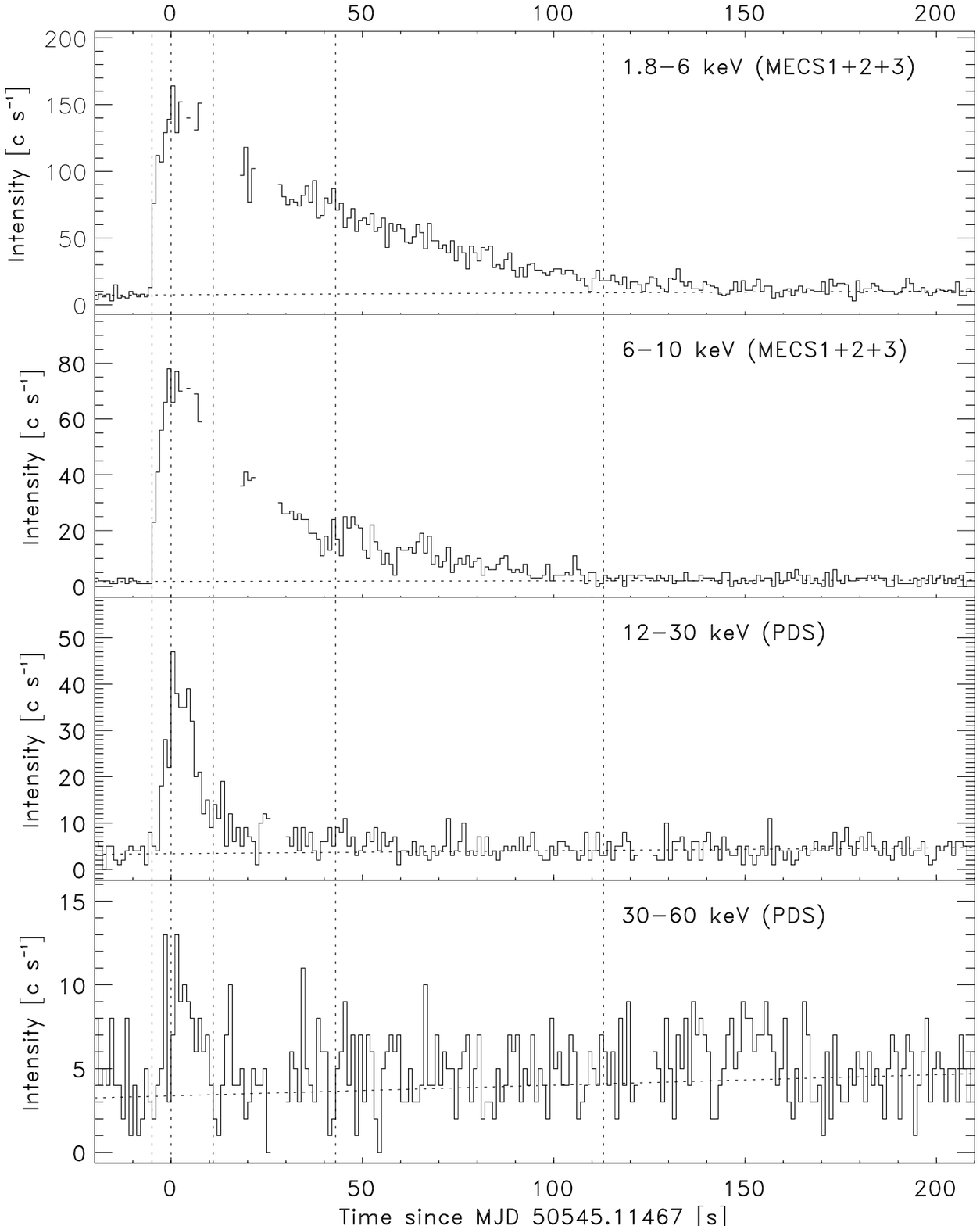,width=\columnwidth,clip=t}

\caption[]{Time profile of the second burst. For an explanation,
see Fig.~\ref{figburstlc1}.
\label{figburstlc2}}
\end{figure}

A power law fit to the 60-150~keV PDS data is 
acceptable ($\chi^2_{\rm r}=0.7$ for 4 dof) and reveals a photon
index of $3.3\pm0.4$ which is close to that found for the 60-300~keV OSSE
data by Strickman et al. (1996) of $3.1\pm0.5$.

The average 0.1 to 200~keV flux is 
$f_{\rm 0.1-200~keV}=(1.93\pm0.10)\times10^{-9}$~\ecs.

We compare the results with those obtained by Del Sordo et al. (1998)
on NFI data taken half a year later on the same source. Del Sordo et al.
find for the black body temperature $0.94\pm0.05$~keV, for the cut off
energy $49\pm3$~keV, for the power law index $1.34\pm0.04$, and for 
$N_{\rm H}\approx4.6\times10^{21}$~cm$^{-2}$. These values are 
consistent with ours. Furthermore, Del Sordo et al. (1998) quote a 2 to 10 
keV flux of $5.6\times10^{-10}$~\ecs\ which is only $\sim$4\% larger than 
what we find.
This indicates that the flux and spectrum of \bron\ did not change 
substantially over half a year.

The optical counterpart is reported to exhibit $E_{B-V}=0.4\pm0.1$ (Motch et 
al. 1994, Barret et al. 1995). Follows that $A_{\rm V}=1.2\pm0.3$ and 
$N_{\rm H}=(2.2\pm0.5)\times10^{21}$~cm$^{-2}$ (according to the conversion of 
$A_{\rm V}$ to $N_{\rm H}$ by Predehl \& Schmitt 1995). An interpolation from 
the HI maps in Dickey \& Lockman (1990) reveals the same value for $N_{\rm H}$. 
This value is inconsistent with the values for the two models of the 
NFI-measured spectrum (Table~\ref{tabpersistentspectrum}). We tried to
accommodate $2.2\times10^{21}$~cm$^{-2}$ with these models. If $N_{\rm H}$
is frozen and the other parameters are left free, the Comptonized model
remains a better description of the data with $\chi^2_{\rm r}=1.498$ 
(133 dof) than the black body plus cut-off power-law model with
$\chi^2_{\rm r}=3.358$ (138 dof). The values of the other parameters in
the Comptonized model are within the error margins as indicated in
Table~\ref{tabpersistentspectrum} except for $kT_{\rm W}$ which is marginally
different at $0.496\pm0.004$~keV. Nevertheless, $\chi^2_{\rm r}=1.498$ is an
unacceptable fit.

\section{The burst emission}
\label{secburst}

Figs.~\ref{figburstlc1} and \ref{figburstlc2} show the time profiles 
of the two bursts in a number of bandpasses from MECS and PDS data at a 
time resolution of 1~s. There are no observations of the bursts with the 
LECS and we omit HP-GSPC data since this instrument has an energy range which
overlaps that of the others. As far as can be judged (there are data 
gaps, probably due to telemetry overflow), the profiles are clean fast-rise 
exponential-decay shapes. The e-folding decay times per bandpass (see 
Table~\ref{tabburstlc}) are identical for both bursts. They are also long 
though not unprecedented, if compared to many other bursters. 
Furthermore, the rise time of the bursts is relatively large (5 to 8~s).

\begin{table}[tb]
\caption[]{Timing parameters and energetics of bursts}
\begin{tabular}{lll}
\hline
                            & burst 1           & burst2 \\
\hline
Peak time (MJD)             & 50544.84838       & 50545.11467 \\
Decay time in 1.8-6~keV (s) & $49.6\pm1.2$      & $49.2\pm0.8$ \\
Decay time in 6-10 keV (s)  & $33.9\pm1.3$      & $34.1\pm0.3$ \\
Bolometric fluence          & $8.1\pm0.5$       & $7.6\pm0.5$\\
\hspace{3mm} ($10^{-7}$~erg~cm$^{-2}$)          &      &\\
\hline
\end{tabular}
\label{tabburstlc}
\end{table}

\begin{figure}[t]
\psfig{figure=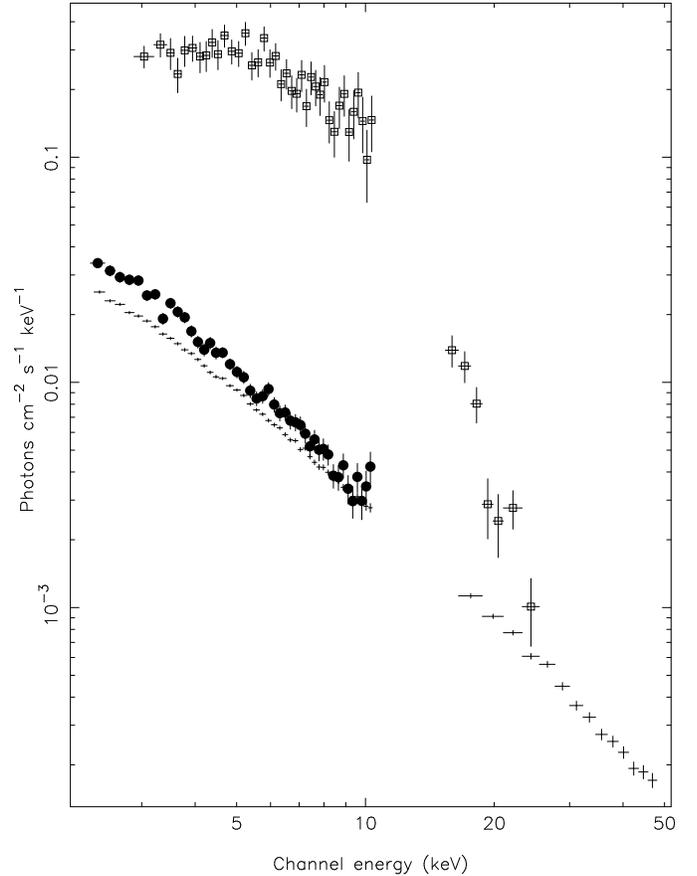,width=\columnwidth,clip=t}

\caption[]{From top to bottom, the photon spectrum of the 0/+11~s
data of the second burst (rectangles), that of the +113/+1113~s 
data of the same burst (filled circles), and that of the non-burst 
data (no symbols, only error bars). Data are from MECS and PDS, except 
for the +113/+1113~s which is MECS only
\label{figburst2spectrum}}
\end{figure}

Each burst was divided in five time intervals (see Figs.~\ref{figburstlc1}
and \ref{figburstlc2}). Relative to the peak time, the intervals are equal
except for the first interval.
The last interval of each burst covers 1000~s to study the slow decay
of the flux to the persistent level. The persistent emission was not subtracted 
in these spectra while the background was. We fitted the MECS spectra in these 
intervals and in the non-burst data with a black body radiation model with
different temperatures plus a power law function whose shape (i.e., photon index) 
is frozen over all intervals. Furthermore, a single level of interstellar plus
circumstellar absorption was fitted to all data through $N_{\rm H}$. PDS 15-30 
keV data were included for the rise and first two decay time intervals of each 
bursts, as well as for the non-burst times up to 50~keV. The fit was 
reasonable with $\chi^2_{\rm r}=1.24$ for 470 dof. Results of the fit are given
in Table~\ref{tabburstfit}. For illustrative purposes, 
a graph is presented in Fig.~\ref{figburst2spectrum} of the photon count rate 
spectra for 2 intervals of the second burst and the non-burst data.

This modeling of the burst spectral evolution shows that
during the brightest parts of the bursts (between 0 and 113~s after the burst
peaks) the black body radius remains constant within an error margin of
roughly 10\% while the temperature decreases from 2 to 1~keV. There is no 
evidence for photospheric expansion.

\begin{table*}[t]
\caption[]{Spectral parameters of two bursts. All data were simultaneously fitted,
including the non-burst data.}
\begin{tabular}{lllllll}
\hline
 & Time interval  & -7/0 s$^{\rm a}$      & 0/11 s     & 11/43 s      & 43/113 s & 113/1113 s \\
\hline
burst 1 & bb $kT$ (keV)  & $1.91\pm0.06$      & $1.96\pm0.05$  & $1.63\pm0.03$  & $1.33\pm0.03$  & $1.10\pm0.06$ \\
  & bb radius (km) & $8.9\pm0.5$        & $10.9\pm0.4$   & $10.8\pm0.4$   & $10.4\pm0.5$   & $3.0\pm0.3$ \\
  & \hspace{5mm} for $d_{\rm 10~kpc}=1$ & & & & & \\
\hline
burst 2 & bb $kT$ (keV)  & $1.80\pm0.08$      & $2.11\pm0.04$  & $1.55\pm0.04$  & $1.36\pm0.03$  & $0.96\pm0.05$ \\
  & bb radius (km) & $9.9\pm0.8$        & $10.7\pm0.3$   & $11.8\pm0.5$   & $9.5\pm0.4$    & $4.2\pm0.4$ \\
  & \hspace{5mm} for $d_{\rm 10~kpc}=1$ & & & & & \\
\hline
\end{tabular}

\noindent
$^{\rm a}$-5/0~s for the second burst
\label{tabburstfit}
\end{table*}

The photon count rate of the black body radiation in the PDS should be 
negligible above 30 keV (i.e., it is about $6\times10^{-3}$ in 30-60 keV 
times that in 12 to 30 keV, for a black body with $kT=2.2$~keV). However, as 
can be seen in Fig.~\ref{figburstlc2}, there is substantial burst emission 
between 30 and 60~keV. In fact, the average 30-60~keV photon count rate of 
the burst in the first 11~s after the burst peak is of order half times that 
in 12-30~keV. This suggests that the burst emission may be Comptonized like 
the persistent emission, although we are not able to verify that spectrally
due to insufficient statistics. 

\section{Discussion} 
\label{secdis}

The thermal nature of the burst spectra with few keV temperatures and 
cooling are typical for a type I X-ray burst (e.g., Lewin et al. 1995, and 
references therein). Such a burst is thought to be due to a thermonuclear
ignition of helium accumulated on the surface of a neutron star. The 
unabsorbed bolometric peak flux of the black body radiation is estimated
at $(2.7\pm0.5)\times10^{-8}$~\ecs. This translates into 
a peak luminosity of $(3.3\pm0.6)\times10^{38}d^2_{\rm 10~kpc}$~\lum. 
Since we do not find evidence for photospheric expansion in the bursts
it is indicated that the burst peak luminosity is below the Eddington 
limit which is $1.8\times10^{38}$~\lum\ for a 1.4~M$_\odot$ neutron star.
Therefore, we expect the distance to be smaller than $7.4\pm0.7$~kpc or 
$\sim8$~kpc. 
Barret et al. (1995) infer from the photometry of the optical 
counterpart that the lower limit to the distance is 4~kpc. The relatively low 
galactic latitude of the source ($-6\fdg1$) does not provide better constraints 
on the distance. For a distance between 4 to 8~kpc, the X-ray luminosity is 
between $3.5\times10^{36}$ and $1.4\times10^{37}$~\lum\ which are fairly 
typical values for LMXB X-ray bursters.

The two bursts presented here show traces of 
Comptonization, like in X~1608--52 
(Nakamura et al. 1989), 1E1724--308 in Terzan 2 (Guainazzi et al. 1998) and 
SAX~J1748.9--2021 in NGC~6440 (In 't Zand et al. 1999). The PDS time profiles 
suggest that the level of Comptonization matches the burst quite 
closely. The time delay is less than a few seconds which suggests that the 
Comptonizing cloud is within $\sim10^{11}$~cm. Unfortunately, this is not a
strong constraint because a 2~h binary orbit implies a system size perhaps one
order of magnitude smaller than that.

Many parameters of the two bursts are equal within narrow error margins: the 
durations within $0.8\pm2.8$\%, the peak temperatures within $7\pm3$\%, 
the peak emission areas within $2\pm5$\%, and the bolometric fluences within 
$6\pm6$\%. This suggests that the physical circumstances for triggering the 
bursts (i.e., the neutron star surface temperature and the composition of 
accreted matter) are the same on the two occasions and, together with the 
prolonged regular bursting and constant persistent flux as measured with WFC, 
testifies to a rather strong stability of the accretion process. This suggests 
a stable accretion disk. In how far this is uncommon among low-luminosity
LMXBs remains to be seen. The knowledge about such LMXBs is as yet incomplete.

The broad-band spectral measurements of the persistent as well as burst emission 
enable a fairly accurate determination of $\alpha$ which is defined as the 
bolometric fluence of the persistent emission between two bursts and that of the 
latter burst. The time between the two bursts is 23,007~s. We are confident that 
no bursts were missed during the data gaps because this time is consistent with 
the quasi-periodicity of the burst recurrence as found from near-simultaneous 
WFC observations with period 5.8~hr and full-width at half maximum of
0.4~hrs (Ubertini et al. 1999). Of all 70 WFC-detected
bursts from \bron, no two were closer to each other than 19,238~s. The fluence of 
the second burst is $(7.6\pm0.5)\times10^{-7}$~erg~cm$^{-2}$. The constant 
persistent emission implies a bolometric fluence between the two bursts of 
$(4.14\pm0.23)\times10^{-5}$~~erg~cm$^{-2}$. Therefore, $\alpha=54\pm5$. This 
confirms the value found from the WFC analysis ($60\pm7$, Ubertini et al. 1999).

\begin{acknowledgements}
We thank the {\em BeppoSAX\/} team at Nuova Telespazio (Rome) for planning and 
carrying out the observation presented here. {\em BeppoSAX\/} is a joint 
Italian and Dutch program.
\end{acknowledgements}

\end{document}